\documentclass[showpacs,10pt,twocolumn,prb]{revtex4-1}

\usepackage{amsmath}
\usepackage{amssymb}
\usepackage{graphicx}
\usepackage{amssymb}
\usepackage{graphics}
\usepackage{epsfig}
\usepackage{CJK}
\usepackage{color}

\begin{document}

\begin{CJK*}{GBK}{}
\title{Magnetotransport properties of MoP$_2$}
\author{Aifeng Wang,$^{1}$ D. Graf,$^{2}$ Yu Liu,$^{1}$ and C. Petrovic$^{1}$}
\affiliation{$^{1}$Condensed Matter Physics and Materials Science Department, Brookhaven National Laboratory, Upton, New York 11973, USA\\
$^{2}$National High Magnetic Field Laboratory, Florida State University, Tallahassee, Florida 32306-4005, USA}

\date{\today}

\begin{abstract}
We report  magnetotransport and de Haas-van Alphen (dHvA) effect studies on MoP$_2$ single crystals, predicted to be type-$\rm{\uppercase\expandafter{\romannumeral2}}$ Weyl semimetal with four pairs of robust Weyl points located below the Fermi level and long Fermi arcs. The temperature dependence of resistivity shows a peak before saturation, which does not move with magnetic field. Large nonsaturating magnetoresistance (MR) was observed, and the field dependence of MR exhibits a crossover from semicalssical weak-field $B^2$ dependence to the high-field linear-field dependence, indicating the presence of Dirac linear energy dispersion. In addition, systematic violation of Kohler's rule was observed, consistent with multiband electronic transport. Strong spin-orbit coupling (SOC) splitting has an effect on dHvA measurements whereas the angular-dependent dHvA orbit frequencies agree well with the calculated Fermi surface. The cyclotron effective mass $\sim$ 1.6$m_e$ indicates the bands might be trivial, possibly since the Weyl points are located below the Fermi level. Interestingly, quasi-two dimensional(2D) band structure is observed even though the crystal structure of MoP$_2$ is not layered.

\end{abstract}
\maketitle
\end{CJK*}

\section{INTRODUCTION}

Weyl fermions in condensed-matter systems have attracted considerable interest and are intensively studied.\cite{Weng, HuangX, XuSY}
Two types of Weyl fermions were found in solid materials. Type-$\rm{\uppercase\expandafter{\romannumeral1}}$ Weyl semimetal (WSM) has an ideal conical Weyl cone in the electronic structure and a closed point-like Fermi surface with Lorentz symmetry, while the type-$\rm{\uppercase\expandafter{\romannumeral2}}$ WSM has a strongly tilted Weyl cone where the Fermi surface consists of an electron and a hole pocket that touch at the Weyl node in a topologically protected manner.\cite{Soluyanov} Thus, Type-$\rm{\uppercase\expandafter{\romannumeral2}}$ WSM could have some exotic properties , such as anisotropic chiral anomaly, anomalous Hall effect, and Klein tunneling.\cite{Soluyanov, Zyuzin, OBrien}

The two originally proposed type-$\rm{\uppercase\expandafter{\romannumeral2}}$ WSM materials are WTe$_2$ and MoTe$_2$.\cite{Soluyanov,Zhijun} However, the arrangement of Weyl points (WPs) is very sensitive to crystal structure. The proximity of WPs with opposite Chern numbers, and the WPs location above the Fermi level makes WPs in WTe$_2$ and MoTe$_2$ unstable and difficult to probe.\cite{Zhijun} WSM can be obtained by either breaking time-reversal symmetry or space-inversion symmetry in a Dirac semimetal.\cite{YanBH}  MoP$_2$, with a similar chemical formula and nonsymmorphic space group Cmc2$_1$ (36), serves as a good candidate for a type-$\rm{\uppercase\expandafter{\romannumeral2}}$ WSM. Recently, MoP$_2$ was predicted to be a type-$\rm{\uppercase\expandafter{\romannumeral2}}$ WSM with stable WPs located below the Fermi level and long Fermi arcs, which can be studied by angle-resolved photoemission spectroscopy (ARPES) measurements.\cite{Autes}

In this paper, we have successfully grown single crystals of MoP$_2$, and performed magnetotransport and de Haas-van Alphen (dHvA) measurements. We found that a peak appears in the temperature dependence of resistivity and that it does not move with field. With increasing magnetic field, magnetoresistance (MR) exhibits a crossover from semiclassical $B^2$ dependence to the linear-field dependence at critical field $B^*$;  the $B^*$ can be described by quadratic behavior for quantum limit with linear energy dispersion. Moreover, a systematic violation of Kohler's rule was observed. The effective mass detected by dHvA is $m \cong 1.6 m_e$, indicating that the band probed by dHvA is relatively trivial, possibly since WPs are located below the Fermi level. Though the crystal structure of MoP$_2$ is three-dimensional (3D) without obvious stacks of layered two-dimensional building blocks, angular-dependent dHvA measurement reveal quasi-2D Fermi surface in MoP$_2$ whereas the results of the dHvA measurement agree well with the calculated Fermi surface.

\section{EXPERIMENTAL DETAILS}

Single crystals of MoP$_2$ were grown by the chemical iodine vapor transport method. Polycrystal of MoP$_2$ was synthesized by heating stoichiometric amounts of Mo and P powders at 500 $^{\circ}$C for 24 h, and then 750 $^{\circ}$C for 48 h. 1 g MoP$_2$ polycrystal was mixed with I$_2$ (15 mg/ml), and then sealed in an evacuated quartz tube. Single crystals were grown in the temperature gradient 1050 $^{\circ}$C (source) to 950 $^{\circ}$C (sink) for two weeks. Small needle-like single crystal with typical size 200 $\mu$m $\times$ 50 $\mu$m $\times$ 20 $\mu$m were obtained. Single crystal x-ray diffraction (XRD) measurements were performed using a Bruker Apex II single crystal x-ray diffractometer with Mo K$_\alpha$ radiation ($\lambda$ = 0.071073 nm) at room temperature. The elemental analysis was performed using an energy-dispersive x-ray spectroscopy (EDX) in a JEOL LSM 6500 scanning electron microscope. Electrical transport were performed using Quantum Design PPMS-9. The dHvA effect at high magnetic field up to 18 T was measured at the National High Magnetic Field Laboratory (NHMFL) in Tallahassee. Resistivity was measured using a standard four contact configuration.

\section{RESULTS AND DISCUSSIONS}

\begin{figure}
\centerline{\includegraphics[scale=0.32]{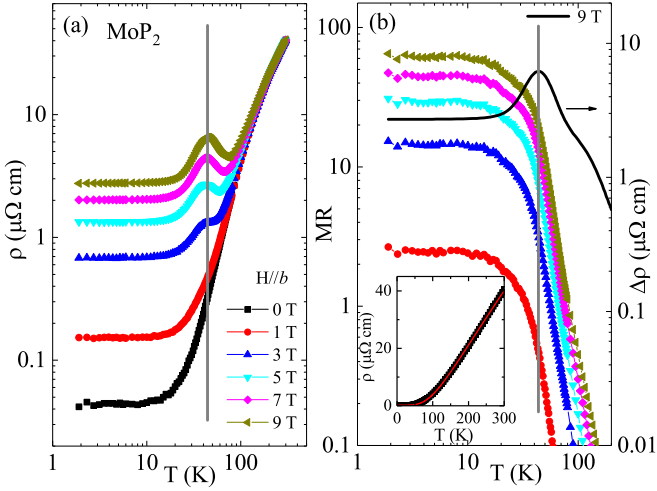}}
\caption{(Color online). (a) Temperature dependence of resistivity in different magnetic fields plotted on a log-log scale, the grey vertical line is a guide to eye. (b) The temperature dependence of the MR = ($\rho$(B) - $\rho$(0T))/$\rho$(0T) at different fields (left),  and the $\Delta$$\rho$ = $\rho$(9T) - $\rho$(0T) (right). (a) and (b) use the same legend. Inset shows $\rho$(0T) fitted with Bloch-Gr\"{u}neisen model.}
\label{res}
\end{figure}

Crystal structure with space group Cmc2$_1$ (36) and lattice parameters $a$ = 3.13 $\rm{\AA}$, $b$ = 11.12 $\rm{\AA}$, and $c$ = 4.94 $\rm{\AA}$ determined by single crystal XRD agree well with previous report.\cite{Rundquist} The average Mo: P atomic ratio determined by EDX is close to 1 : 2, consistent with the composition of MoP$_2$.

The temperature dependence of resistivity in different magnetic fields were measured with $B$ parallel to the crystallographic $b$ axis and current parallel to the $a$ axis, as shown in Fig. 1(a). $\rho (0T)$ shows a metallic behavior with $\rho$(2K) = 0.042 $\mu\Omega$cm and RRR = 965, indicating the high quality of our single crystals. $\rho (0T)$ between 2 K to 300 K can be well fitted by the Bloch-Gr\"{u}neisen (BG) model:\cite{Ziman}
\[
\rho (T)=\rho _0 + C(\frac{T}{\Theta_D})^5\int_{0}^{\frac{\Theta_D}{T}}\frac{x^5}{(e^x-1)(1-e^{-x})}dx
\]
where $\rho_0$ is the residual resistivity and $\Theta_D$ is the Debye temperature. The fitting gives $\Theta_D$ = 591 K, similar to that of WP$_2$. The good BG model fit suggests that the phonon scattering dominates in the absence of magnetic field. As shown in Fig. 1(a), a magnetic field-induced resistivity upturn was observed at low temperature, in contrast to other semimetals such as TaAs, WTe$_2$, and LaSb,\cite{HuangX, Cava, Tafti} the resistivity diminishes with further decrease in temperature and saturates below $\sim$ 14 K. Tis peak in temperature-dependent resistivity at 45 K which does not move with field, is shown by the vertical line in Fig. 1(a). We note that a weak peak was also observed in WP$_2$.\cite{Kumar} A peak in temperature dependence of MR in pure aluminum and indium decreases monotonically with the increase in magnetic field, which can be explained by the two band model.\cite{Snodgrass, Huffman} Even though the peak can be observed in $\Delta\rho$, the MR monotonously increases with decreasing temperature and saturates at $\sim$ 14 K, as shown in Fig. 1(b). This is similar to copper.\cite{Huffman, Schwartz} In order to understand the peak effect we measured magnetic field dependence of MR at different temperatures.

\begin{figure}
\centerline{\includegraphics[scale=0.3]{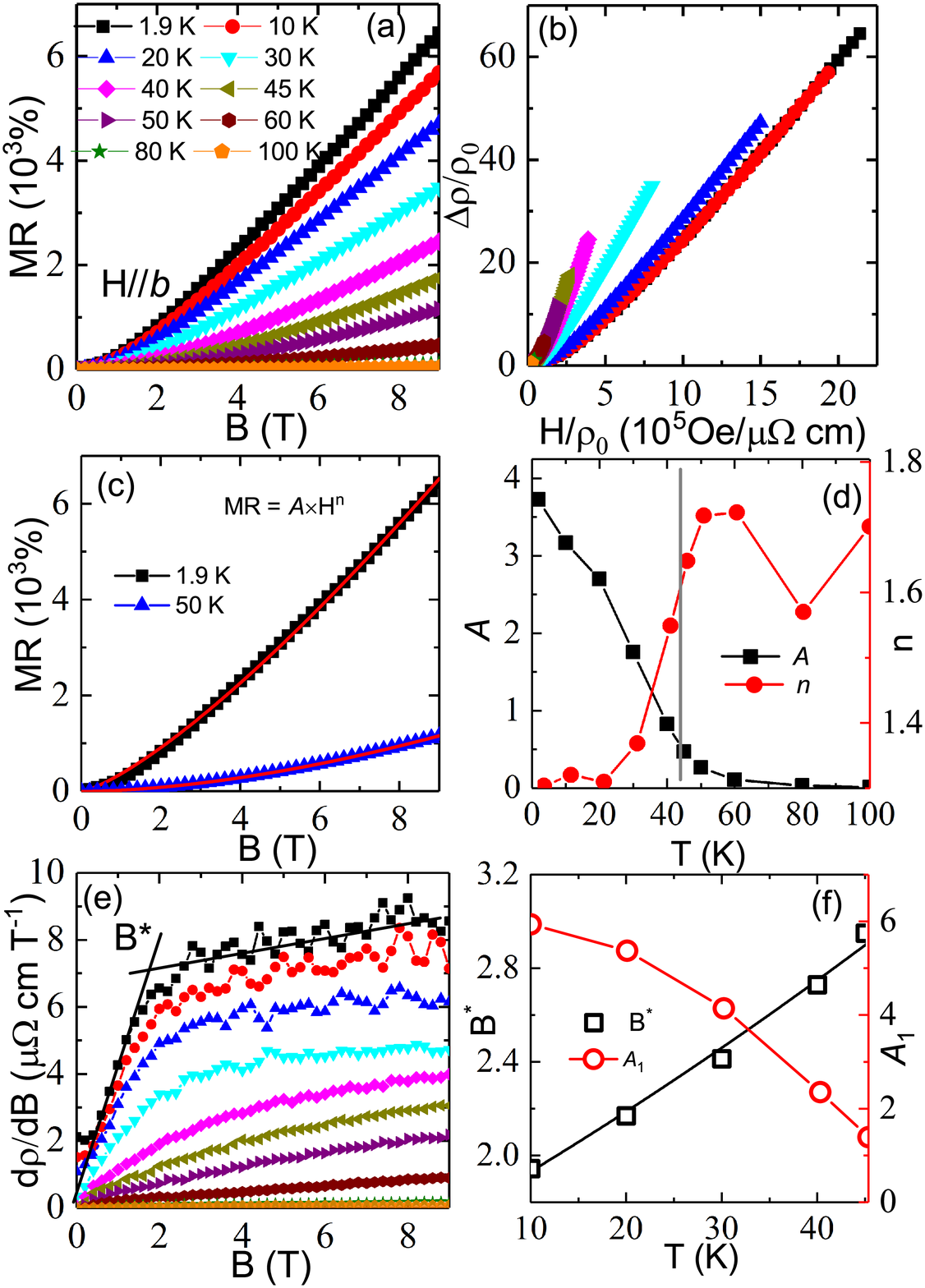}}
\caption{(Color online). The MR versus magnetic field at different temperatures for MoP$_2$ with $H$//$b$. (b) A Kohler plot for MoP$_2$. (c) Typical magnetic field dependence of MR curves, the solid lines are fits using MR = $A\times$$H^n$. (d) Temperature dependence of $A$ (left) and $n$ (right) in MR = $A\times$$H^n$. (e) The field derivative of MR ($d$(MR)/$d B$ at different temperatures, solid lines show the criterion used to determine the critical field $B^*$. (f) Temperature dependence of the critical field $B^*$ (left); the black solid line is the fitting of $B^*$ by $B^* = \frac{1}{2e\hbar\upsilon_F^2}(E_F+k_BT)^2$. The red circle corresponds to high-field MR linear coefficient $A_1$.}
\label{magnetism}
\end{figure}

As shown in Fig.2(a), square to linear transition of field dependent MR was observed, and the MR can reach as high as 6,450\% at 1.9 K and 9 T. This is similar to Ta$_3$S$_2$,\cite{ChenD} one or two order of magnitude smaller when compared to WTe$_2$ and LaSb.\cite{Cava, Tafti} According to semiclassical transport theory, if there is a single type of charge carrier and scatting time in a metal, Kohler's rule states that the relative change in resistivity $\Delta\rho$/$\rho_0$ in a magnetic field $H$ is a universal function of $H/\rho_0$, where $\rho_0$ is the zero-field resistivity at certain temperature.\cite{Pippard} As shown in Fig. 2(b), while the data below 45 K deviate from Kohler's rule, the data above 45 K still fall on same curve, suggesting that the transport above 45 K  is dominated by a single scattering process. Multiple reasons can lead to the breakdown of Kohler's law, possibly due to a multiband effect with different scattering times here. We try to fit the MR with a simple power law MR = $A$ $\times$ $H^n$, the typical data is shown in fig. 2(c). The MR above 45 K can be well fitted in whole range, but there are some deviations in the low field range of the data below 45 K. The obtained parameters are shown in Fig. 2(d), $A$ monotonously decreases with temperature increase and $n$ increases with increasing temperature and saturates at $\sim$ 1.7. $n$ should be 2 in a conventional metal, while linear MR is often observed in Dirac semimetals. One of the possible explanations for the temperature dependence of $n$ is that the Dirac band plays a more important role as the temperature is lowered. In order to characterize the crossover behavior from weak-field $B^2$ dependence to the high-field linear dependence, we present the field derivative of the MR, $d$MR/$d$B, in Fig. 2(e). Linear field dependence for $d$MR/$d$B  in low fields agree with the semiclassical MR $\sim$ $A_2$$B^2$. With field increasing, $d$MR/$d$B reduces from linear behavior to a weak filed dependence saturating behavior above a characteristic field $B^*$, indicating that MR is dominated by a linear field dependence plus a small quadratic term [$\Delta\rho/\rho = A_1B + o(B^2)$] at high field region.\cite{Kefeng, Fisher} Since the splitting between the lowest and first landau level (LL) for Dirac state is described as $\Delta_{LL} = \pm\upsilon_F\sqrt{2e{\hbar}B}$, while $\Delta_{LL} = \frac{e{\hbar}B}{m^*}$ for a conventional parabolic band, quantum limit is easily reached for Dirac band.\cite{Kefeng} As shown in Fig. 2(f), $B^*$ can be described by critical field for quantum limit at a specific temperature $B^* = \frac{1}{2e\hbar\upsilon_F^2}(E_F+k_BT)^2$,\cite{Huynh} suggesting the linear MR likely originates from the Dirac states.

\begin{figure}
\centerline{\includegraphics[scale=0.3]{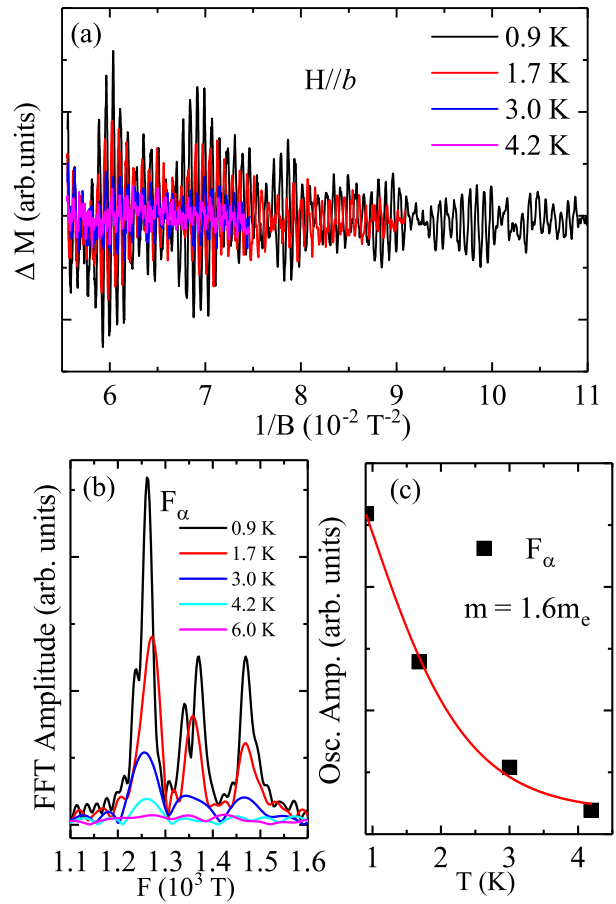}}
\caption{(Color online). (a) dHvA oscillatory components at different temperatures obtained by subtracting smooth background. (b) FFT spectra for the dHvA oscillations in (a). (c) Temperature dependence of the oscillating amplitude at $F_{\alpha}$ = 1261 T, solid line is fitted using Lifshitz-Kosevich formula.}
\label{Hall}
\end{figure}

Quantum oscillation is a powerful tool in investigating Weyl/Dirac materials. We measured de Haas-van Alphen (dHvA) oscillations in MoP$_2$ at different temperatures, as shown in Fig.3(a). Beat patterns were observed, indicating that multiple frequencies contribute to the oscillations. The corresponding fast Fourier transform (FFT) spectra is shown in Fig. 3(b), similar to that of WP$_2$. Three frequencies are observed: 1261 T, 1371 T, and 1469 T. According to Onsager relation, $F = (\Phi_0/2\pi^2)A_F$, where $\Phi_0$ is the flux quantum and $A_F$ is the orthogonal cross-sectional area of the Fermi surface, the Fermi surface is estimated to be 12 nm$^{-2}$, 13 nm$^{-2}$, and 14 nm$^{-2}$, corresponding to 4.7\%, 5.1\%, and 5.5\% of the total area of the Brillouin zone in the $ac$ plane. The oscillations dampen quickly with temperature and disappear above 4.2 K, indicating a heavy cyclotron mass in MoP$_2$. The temperature dependence of amplitude of the oscillations is fitted using the Lifshitz-Kosevich formula,\cite{Shoeneberg} $A$$\sim$[$\alpha$$m^{*}$(T/B)/$\sinh$($\alpha$ $m^{*}$T/B)] where $\alpha$ = 2$\pi$$^2$$k_{\rm B}$/e$\hbar$ $\approx$ 14.69 T/K, $m^{*}$ = $m$/$m_{e}$ is the cyclotron mass ratio ($m_{e}$ is the mass of free electron), as exhibited in Fig. 3(c). The fitting gives $m^{*}$ $\cong$ 1.6, similar to that in WP$_2$, possibly due to electron-phonon many body interactions.\cite{Kumar} The effective cyclotron mass suggests that the band detected by the dHvA effect might be trivial, possibly since the Weyl points are located below the Fermi level.\cite{Autes} The trivial Fermi surfaces are not consistent with the Dirac states indicated by linear MR. A possible explanation is  that the Dirac state usually leads to a small Fermi surface and therefore might can not been detected by dHvA measurement. This is similar to the case in MoP, where a triply degenerate point with Dirac-like disperson well below Fermi level have been discovered.\cite{LvBQ}

\begin{figure}
\centerline{\includegraphics[scale=0.32]{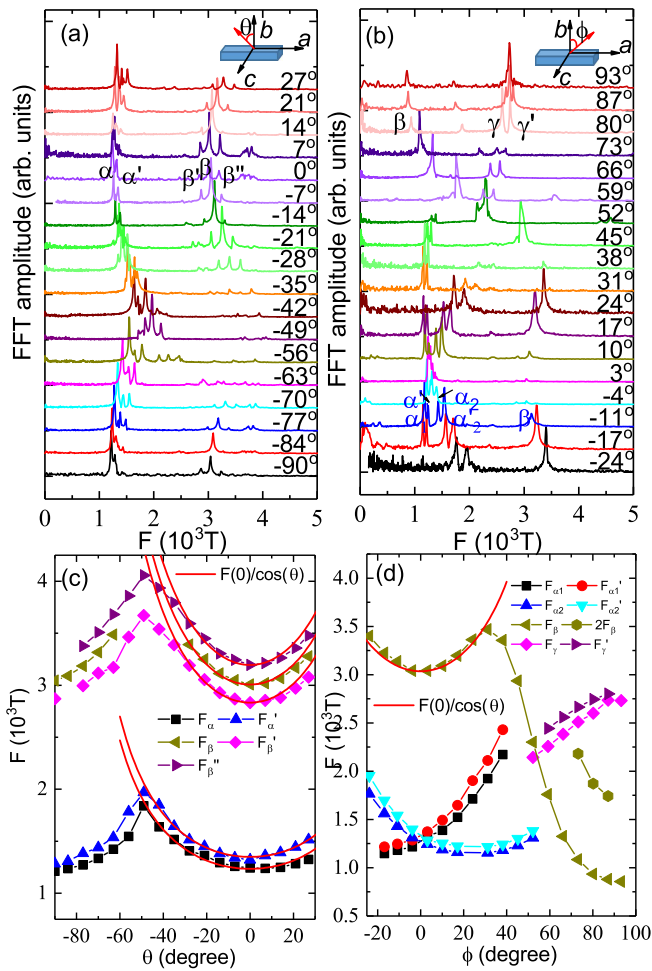}}
\caption{(Color online). FFT spectra of dHvA oscillaiton with field rotated in the $bc$ plane (a) and in the $ab$ plane (b); the spectra are nomalized and shifted vertically for clearity. (c,d) Angular dependence of the oscillation frequency corresponding to the oscillation in (a) and (b), respectively. Red solid lines are fits with 2D model $F(0)$/cos($\theta$). }
\label{Hall}
\end{figure}

Angular-dependent dHvA oscillations provide further insights into Fermi surface properties.The FFT spectra of dHvA oscillations are shown in Fig. 4(a,b) for magnetic field rotated in the $bc$ and $ab$ planes, respectively. The positions of the FFT peaks are summarized in Fig. 4(c) and Fig. 4(d). The FFT peaks can be generally divided into two groups, one group around 1200 T, and another group around 3000 T. Assuming MoP$_2$ have similar band structure as WP$_2$,\cite{Autes,Kumar} lower frequencies around 1200 T should arise from the oscillation of spaghetti-like open hole Fermi surface, while higher frequencies around 3000 T are due to the oscillation of bow-tie-like closed electron Fermi surface. Split peaks and small peaks around the main peaks might due to strong spin-orbit coupling.\cite{Kumar} Even though the crystal structure is three-dimensional and quite different from the quasi-2D structure of SrMnBi$_{2}$ with square Bi layers,\cite{Kefeng} all peaks show  quasi-2D angle dependence [$F(0)/cos(\theta)$] at low angles around H//$b$, indicating quasi-2D FS in MoP$_2$ [Fig. 4(c)]. When the magnetic field is rotated in the $ab$ plane, peaks corresponding to hole pocket show behavior of two elliptical FS enlongated $\sim$ 25$^{\rm o}$ away from $b$ axis. This is consistent with the calculated spaghetti-like hole pocket.\cite{Autes,Kumar} The spaghetti-like hole pocket is flat along the $ab$ plane, indicating quasi-2D like behavior with magnetic field rotated in $bc$ plane, while it is rather warp in the $bc$ plane with two hole pockets bent at opposite directions. This gives rise to the behavior in Fig. 4(d). F$_{\beta}$ also shows quasi-2D angle dependence with field rotated in $ab$ plane with $\phi$ $\leq$ 30$^{\rm o}$, however, with further increase the angle, F$_{\beta}$ decreases quickly to 856 T at $\phi$ = 90$^{\rm o}$ (H//$a$). This can be explained by the bow-tie-like electron pocket, when $\phi$ $\leq$ 30$^{\rm o}$; quantum oscillations are due to the the orbit cross all the pocket, flat side wall gives rise to quasi-2D behavior. When $\phi$ $>$ 30$^{\rm o}$ quantum oscillations are due to the orbit on the neck of the bow tie, which decreases quickly with angle and show minimum value with $\phi$ = 90$^{\rm o}$ (H//$a$), even smaller than hole pocket.\cite{Autes,Kumar} The observation of strong spin-orbit coupling induced band splitting indicates that spin-orbit coupling and the related spin and orbital angular momentum textures might play an important role in large MR in MoP$_2$, similar to that in WTe$_2$.\cite{JiangJ}

\section{CONCLUSIONS}

In conclusion, our MR studies confirm the presence of both Dirac states and parabolic band with enhanced quasiparticle mass in MoP$_{2}$. The dHvA measurements reveal quasi-2D multiband transport. Whereas the angular dependence of dHvA is consistent with the calculated Fermi surface, the splitting of the bands suggests strong spin-orbit coupling.

\section*{Acknowledgements}

We thank David Szalda for help with Bruker APEXII measurements and John Warren for his help with SEM measurements. Work at BNL was supported by the U.S. DOE-BES, Division of Materials Science and Engineering, under Contract No. DE-SC0012704. Work at the National High Magnetic Field Laboratory is supported by the NSF Cooperative Agreement No. DMR-1157490, and by the state of Florida.

\end{document}